\documentclass[manuscript]{acmart}
\usepackage{makecell}
\usepackage{multirow}
\AtBeginDocument{%
  \providecommand\BibTeX{{%
    \normalfont B\kern-0.5em{\scshape i\kern-0.25em b}\kern-0.8em\TeX}}}

\copyrightyear{2023}
\acmYear{2023}
\setcopyright{rightsretained}
\acmConference[RecSys '23]{Seventeenth ACM Conference on Recommender Systems}{September 18--22, 2023}{Singapore, Singapore}
\acmBooktitle{Seventeenth ACM Conference on Recommender Systems (RecSys '23), September 18--22, 2023, Singapore, Singapore}\acmDOI{10.1145/3604915.3608874}
\acmISBN{979-8-4007-0241-9/23/09}

\begin{document}

\title{Heterogeneous Knowledge Fusion: A Novel Approach for Personalized Recommendation via LLM}

\author{Bin Yin}
\authornotemark[1]
\email{yinbin05@meituan.com}
\author{JunJie Xie}
\authornote{Both authors contributed equally to this research.}
\email{xiejunjie02@meituan.com}
\author{Yu Qin}
\email{qinyu12@meituan.com}
\author{ZiXiang Ding}
\email{dingzixiang@meituan.com}
\author{ZhiChao Feng}
\email{fengzhichao03@meituan.com}
\affiliation{%
  \institution{Meituan}
  \city{Beijing}
  \country{China}
}

\author{Xiang Li}
\authornote{Corresponding author.}
\email{leo.lx007@qq.com}
\author{Wei Lin}
\email{lwsaviola@163.com}
\affiliation{%
  \institution{Unaffiliated}
  \city{Beijing}
  \country{China}}

\renewcommand{\shortauthors}{Yin and Xie, et al.}

\begin{abstract}
The analysis and mining of user heterogeneous behavior are of paramount importance in recommendation systems. 
However, the conventional approach of incorporating various types of heterogeneous behavior into recommendation models leads to feature sparsity and knowledge fragmentation issues. 
To address this challenge, we propose a novel approach for personalized recommendation via Large Language Model (LLM), by extracting and fusing heterogeneous knowledge from user heterogeneous behavior information. 
In addition, by combining heterogeneous knowledge and recommendation tasks, instruction tuning is performed on LLM for personalized recommendations.
The experimental results demonstrate that our method can effectively integrate user heterogeneous behavior and significantly improve recommendation performance.
\end{abstract}




\received{1 June 2023}
\received[revised]{29 June 2023}
\received[accepted]{27 July 2023}

\maketitle

\section{Introduction}

The analysis and mining of user behavior is a crucial aspect in recommendation systems.
In the context of Meituan Waimai, user behavior exhibits heterogeneous characteristics, including various behavior subjects, content, scenarios.
The current industry approach mostly involves continuously adding various heterogeneous behavior to the traditional recommendation models, which brings two obvious problems.  
Firstly, the multitude of behavior subjects leads to sparse features that pose challenges to efficient modeling.  
Secondly, separating the modeling of user, merchant, and commodity behavior ignores the fusion of heterogeneous knowledge among behavior.
However, we have noticed that heterogeneous user behavior contain rich semantic knowledge, and using semantics to represent and reason about user behavior can more effectively promote heterogeneous knowledge fusion and capture user interests.
\begin{figure}[h]
  \centering
  \includegraphics[width=\linewidth]{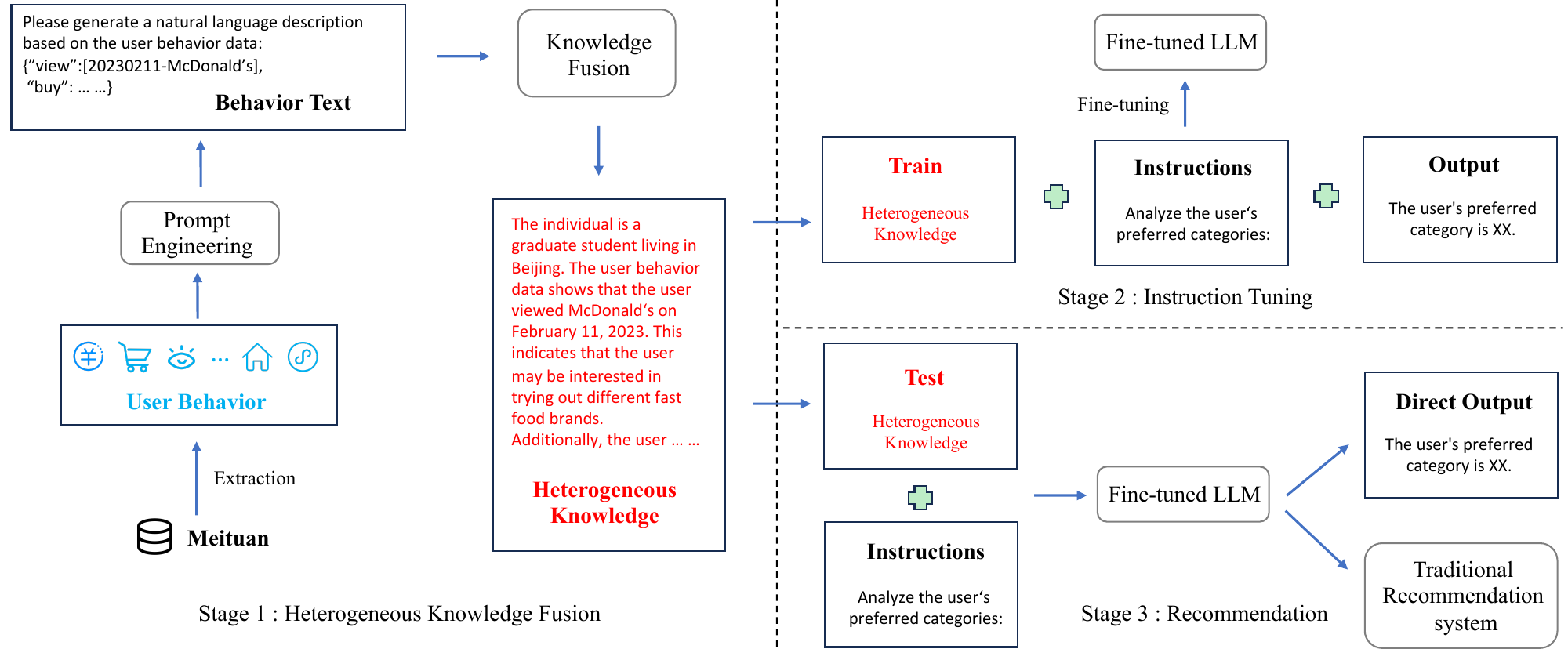}
  \caption{The overall framework of HKFR, including heterogeneous knowledge fusion (left), fine-tuning and recommendation (right).}
  \Description{HKFR model}
  \label{HKFR}
\end{figure}

LLMs have shown remarkable capabilities in various fields, thanks to rich semantic knowledge and powerful inferential reasoning \cite{chowdhery2022palm,zhao2023survey}. 
We have designed a new user behavior modeling framework via LLM, which extracts and integrates heterogeneous knowledge from heterogeneous behavior information of users, and transforms structured user behavior into unstructured heterogeneous knowledge.
In the field of recommendation, there have been some attempts to use LLM for personalized recommendation. Liu et al \cite{liu2023chatgpt} have used LLM to express recommendation tasks in natural language based on context cues, but only adopt a single user behavior modeling, ignoring the fusion of user heterogeneous knowledge.
In addition, there are differences between the training task of LLMs and the recommendation task, which requires fine-tuning for specific recommendation tasks.
Therefore, we propose \underline{\textbf{H}}eterogeneous \underline{\textbf{K}}nowledge \underline{\textbf{F}}usion, a new approach of personalized \underline{\textbf{R}}ecommendation via LLM (\textbf{HKFR}).
Firstly, we design a new user behavior modeling framework using LLM, which extracts and integrates heterogeneous knowledge from heterogeneous behavior information of users.
Then we combine heterogeneous knowledge and design instruction datasets for recommendation tasks to fine-tune LLM.
Finally, using the constructed recommendation task instruction and heterogeneous knowledge as input and the fine-tuned LLM for calculation, we output the recommended features and results for users.

In summary, our contributions are as follows:
\begin{itemize}
\item We propose a new paradigm for user modeling, which extracts and integrates heterogeneous knowledge from heterogeneous user behavior information, transforming structured user behavior into unstructured heterogeneous knowledge, effectively capturing user interests.
\item We construct a personalized recommendation model based on LLM, combining recommendation tasks and heterogeneous knowledge, fine-tuning LLM to make it more effective for tasks in recommendation scenarios.
\item We have conducted sufficient offline and online experiments on Waimai dataset, verifying that our approach can fully integrate heterogeneous user behavior and effectively improve recommendation performance.
\end{itemize}

\section{Methodology}
In this section, we will introduce the framework of HKFR, which includes three stages as shown in Figure \ref{HKFR}.
\subsection{Heterogeneous Knowledge Fusion}
In the stage of heterogeneous knowledge fusion, we utilize the rich semantic knowledge and powerful reasoning abilities of LLM to facilitate the integration of heterogeneous knowledge.

In the context of Meituan Waimai, user heterogeneous behavior includes: multiple behavior subjects, such as merchants and products; multiple behavior contents, such as exposure, clicks, and orders; multiple behavior scenarios, such as APP homepage and mini-programs.
To address diverse user heterogeneous behavior, we extract and structure user behavior data from the database with users as the core.
Then, in the prompt engineering module, we design and construct different structured templates for diverse user behavior, expressing heterogeneous behavior as templated text language.
Next, in the knowledge fusion module, we employ ChatGPT \cite{OpenAI2023GPT4} to perform heterogeneous knowledge fusion on the behavior text, obtaining heterogeneous knowledge text.
The heterogeneous knowledge generated based on user behavior will be used for the fine-tuning and recommendation stages of LLM.
\subsection{Instruction Tuning}
The fine-tuning process aims to help LLM better understand heterogeneous knowledge and further improve its accuracy and adaptability in recommendation tasks \cite{ouyang2022training}.
We construct an instruction dataset based on the recommendation task and heterogeneous knowledge, which includes input, instruction, and output.
The input is the heterogeneous knowledge generated in the Section 2.1.
The instruction and output are a series of task descriptions and target results generated specifically for recommendation. The instruction includes recommendations for user preferences on categories, price, and merchants, among others, while the output is the true label of the user's next order.
Based on the constructed instruction dataset, we perform instruction tuning on LLM. Specifically, we choose the open-source model ChatGLM-6B \cite{du2022glm} as our base LLM and adopt the Lora method for fine-tuning \cite{hu2021lora}.
\subsection{Recommendation}
Given a user, retrieve user behavior heterogeneous knowledge from the database as input for LLM. Then, design instruction based on the recommendation task, perform reasoning and calculations, and output the recommended results for the user.
These instructions include predicting user preferences for categories, prices, and other features, as well as their next click on a merchant.
The predicted results can be output as direct recommendations in natural language form and can also be used as semantic features to enhance the recommendation effect by concatenating with the existing features in traditional recommendation models.
\section{experiment}
In order to evaluate the performance of the model, we select two different recommendation tasks: recommending categories and POIs (points of interest) to user.
\subsection{Implementation}
For offline experiment, we select the dataset from March to April 2023 of Meituan Waimai, design 20 recommendation task instructions, and construct a total of 100,000 users and 1 million instruction data. The testing set is selected from the samples on May 9th, 2023, consisting of 10,000 instruction data with two tasks, recommending POIs and categories.
Each data includes user statistical features, POI features, user heterogeneous behavior sequence features, label, etc.
Due to the limitation of the input length, the user sequence length is limited to 300. Additionally, user and POI data are anonymized before being inputted into LLM.
To evaluate the recommendation effectiveness, we select top-k HR and top-k NDCG, with k={5, 10}.
To demonstrate the effectiveness of our method, we compare it with traditional recommendation methods Caser \cite{tang2018personalized}, BERT4Rec \cite{sun2019bert4rec}, and language models P5 \cite{geng2022recommendation}, ChatGLM-6B \cite{du2022glm}.
\subsection{Results and Anaysis}
Table \ref{Result} displays the experimental results, which demonstrate that our model outperforms multiple baselines on the Waimai dataset with significant improvements.
\begin{table*}[]
\renewcommand{\arraystretch}{1.2}
\caption{Experimental results on two recommendation tasks, with bold indicating the best results and italic indicating ablation experiments."no-HKF" means removing heterogeneous knowledge fusion, and "no-IT" means removing instruction tuning.}
\label{Result}
\begin{tabular}{c|cccc|cccc}
\hline
\multirow{2}{*}{\makecell{Methods}} & \multicolumn{4}{c|}{Category}                                         & \multicolumn{4}{c}{POIs}                                         \\ \cline{2-9} 
                         & HR@5            & NDCG@5          & HR@10           & NDCG@10         & HR@5            & NDCG@5          & HR@10           & NDCG@10         \\ \hline
Caser                    & 0.1152          & 0.1063          & 0.2147          & 0.1320          & 0.0897          & 0.0770          & 0.1842          & 0.1012          \\
BERT4Rec                 & 0.1217          & 0.1140          & 0.2196          & 0.1440          & 0.0875          & 0.0744          & 0.1811          & 0.0995          \\
P5                       & 0.1416          & 0.1384          & 0.2477          & 0.1589          & 0.1218          & 0.1159          & 0.2187          & 0.1260          \\
ChatGLM-6B               & 0.1074          & 0.1019          & 0.2038          & 0.1254          & 0.0785          & 0.0720          & 0.1702          & 0.0872          \\ \hline
$HKFR_{no-IT}$              & 0.1241          & 0.1175          & 0.2267          & 0.1415          & 0.1014          & 0.0952          & 0.2050          & 0.1165          \\
$HKFR_{no-HKF}$             & 0.1813          & 0.1308          & 0.2825          & 0.1580          & 0.1421          & 0.0975          & 0.2432          & 0.1270          \\
\textbf{HKFR}           & \textbf{0.2160} & \textbf{0.1586} & \textbf{0.3007} & \textbf{0.1840} & \textbf{0.1726} & \textbf{0.1243} & \textbf{0.2610} & \textbf{0.1525} \\ \hline
\end{tabular}
\end{table*}
Compared with traditional sequential recommendation methods such as Caser and BERT4Rec, our model demonstrates superior performance. 
This is mainly attributed to the effective fusion of heterogeneous knowledge by LLM.
Our approach also achieves the best performance compared to language models such as P5 and ChatGLM. 
This is mainly due to the fine-tuning of professional datasets, which effectively improves the mismatch between LLM and recommendation tasks.

We conduct several experiments to validate the effectiveness of key modules in our HKFR. Compared to $HKFR_{no-HKF}$, which removes the heterogeneous knowledge fusion stage, HKFR demonstrates superior performance. This is mainly attributed to the accurate capture of user interests by fusing heterogeneous knowledge. Compared to $HKFR_{no-IT}$, which removes the instruction tuning stage, HKFR achieves better results, indicating that instruction tuning can effectively promote LLM to adapt to downstream recommendation tasks.

HKFR was applied to the Meituan Waimai recommendation system for online A/B testing, by utilizing the computed features from the previous day's search query of users and inputting them into the real-time computation on the current day. The experiment ran for a period of May 9, 2023 to May 19, 2023.
The results indicate that HKFR achieved improvements of 2.45\% in CTR and 3.61\% in GMV for cold start users, while there was no significant effect found on other users. This is attributed to the insufficient catering expertise of LLM, which makes it challenging to fully comprehend and integrate heterogeneous behavior. Further training of LLM in the catering domain is necessary to address this limitation.
\section{conclusion}
In this article, we propose HKFR,  a new approach of personalized recommendation via LLM, which is based on the fusion of heterogeneous knowledge.
By leveraging the two stages of heterogeneous knowledge fusion and instruction tuning, HKFR can effectively model user heterogeneous behavior. 
Extensive experiments on the Waimai dataset have validated the ability of HKFR to improve recommendation performance.
In the future, we will focus on further training HKFR in the catering domain to better integrate heterogeneous knowledge and enhance recommendation performance.
\bibliographystyle{ACM-Reference-Format}
\bibliography{sample-base}

\end{document}